# Session Initiation Protocol Attacks and Challenges

Hassan Keshavarz [1], Mohammad Reza Jabbarpour Sattari [2] and Rafidah Md Noor [3]

[1, 2, 3] Faculty of Computer Science and Information Technology, University of Malaya, Kuala Lumpur, 50603, Malaysia

**Abstract.** In recent years, Session Initiation Protocol (SIP) has become widely used in current internet protocols. It is a text-based protocol much like Hyper Text Transport Protocol (HTTP) and Simple Mail Transport Protocol (SMTP). SIP is a strong enough signaling protocol on the internet for establishing, maintaining, and terminating session. In this paper the areas of security and attacks in SIP are discussed. We consider attacks from diverse related perspectives. The authentication schemes are compared, the representative existing solutions are highlighted, and several remaining research challenges are identified. Finally, the taxonomy of SIP threat will be presented.

**Keywords:** Cryptography, Cryptanalysis, SIP.

## 1. Introduction

In 2002, the Internet Engineering Task Force (IETF) proposed the Session Initiation Protocol (SIP) as the IP-based telephony protocol [1]. SIP, as defined in RFC 3261, was chosen by Third-Generation Partnership Project (3GPP) as the protocol for multimedia applications in 3G mobile networks. Internet protocols such as Hyper Text Transport Protocol (HTTP) and Simple Mail Transport Protocol (SMTP) use SIP because it is a text-based signaling protocol [2] and a protocol such as H.323 is a lightweight and flexible signaling. SIP is a signaling protocol for establishing, maintaining, and terminating a multimedia session with one or more participants. It is based on the application layer and as a result, it is a request-response protocol which makes a request for the server and sends a response to the client. However, SIP specifications do not provide strong security mechanisms because it functions based on HTTP Digest authentication, noted in RFC 2617 [3].

The remainder of the paper is organized as follows: Section 2 is an overview of the SIP architecture and authentication procedure. Section 3 discusses the weaknesses of the original SIP authentication scheme and scheme improvements from various research works. Section 4 discusses a security model related to Canetti-Krawczyk. Section 5 discusses the performance comparison between reviewed security models. Section 6 concludes the paper.

## 2. SIP Architecture and Authentication Procedure

This section introduces SIP architecture and authentication procedure. The SIP is composed of a proxy server, a user agent, redirect server, a register server and a location server. The function of each component is described as follows:

User Agent: a logical entity such as a caller (the user agent client (UAC)) or a callee (the user agent server (UAS)).

Proxy Server: A proxy server forwards a request and response between a callee and caller. When the proxy server receives a request, it forwards it to the current callee’s location and then forwards the response

[+] Corresponding author. Tel.: +60172950923 .
*E-mail address*: keshavarz_hassan@ieee.org.

from the callee to the caller.

The operational steps in the proxy mode require a two-way call communication listed below:

- The proxy server accepts the INVITE request from the client.
- The proxy server contacts the location server to request the called party UA's address.
- The location server identifies the called party's location and provides the address of the target server.
- The INVITE request is forwarded to the address of the location that is returned.
- The called party UA alerts the user. The user answers the call.
- The UA returns a 200 OK indication to the requesting proxy server.
- The 200 OK requests are forwarded from the proxy server to the calling party UA.
- The calling party UA confirms receipt of the 200 OK by issuing an ACK request, which is send to the proxy or directly to the called party UA.
- The proxy forwards the ACK to the called party UA.

Figure 1 illustrates the communication exchange for the INVITE method using the proxy server.

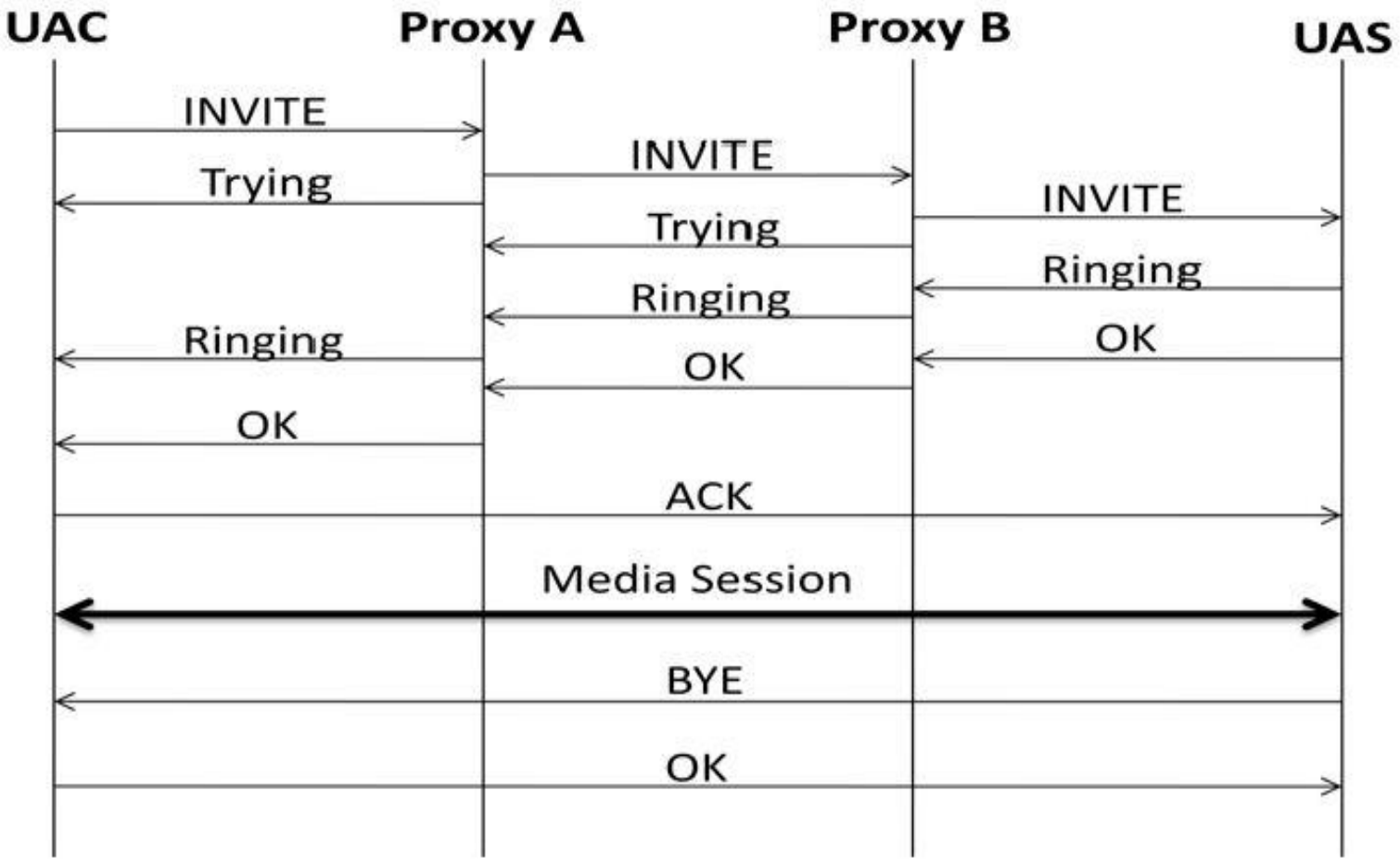

Fig. 1: Proxy model of operation

Redirect Server: a UAS that generates 300 class SIP responses to the requests it receives, directing the UAC to contact an alternate set of Uniform Resource Identifiers (URI).

Location Server: The responsibility of the location server is to maintain information on the user agent's current location. It also services the proxy server, redirects it, and registers the server for them to look up or register the user agent's location.

The SIP architecture is demonstrated in the figure 2.

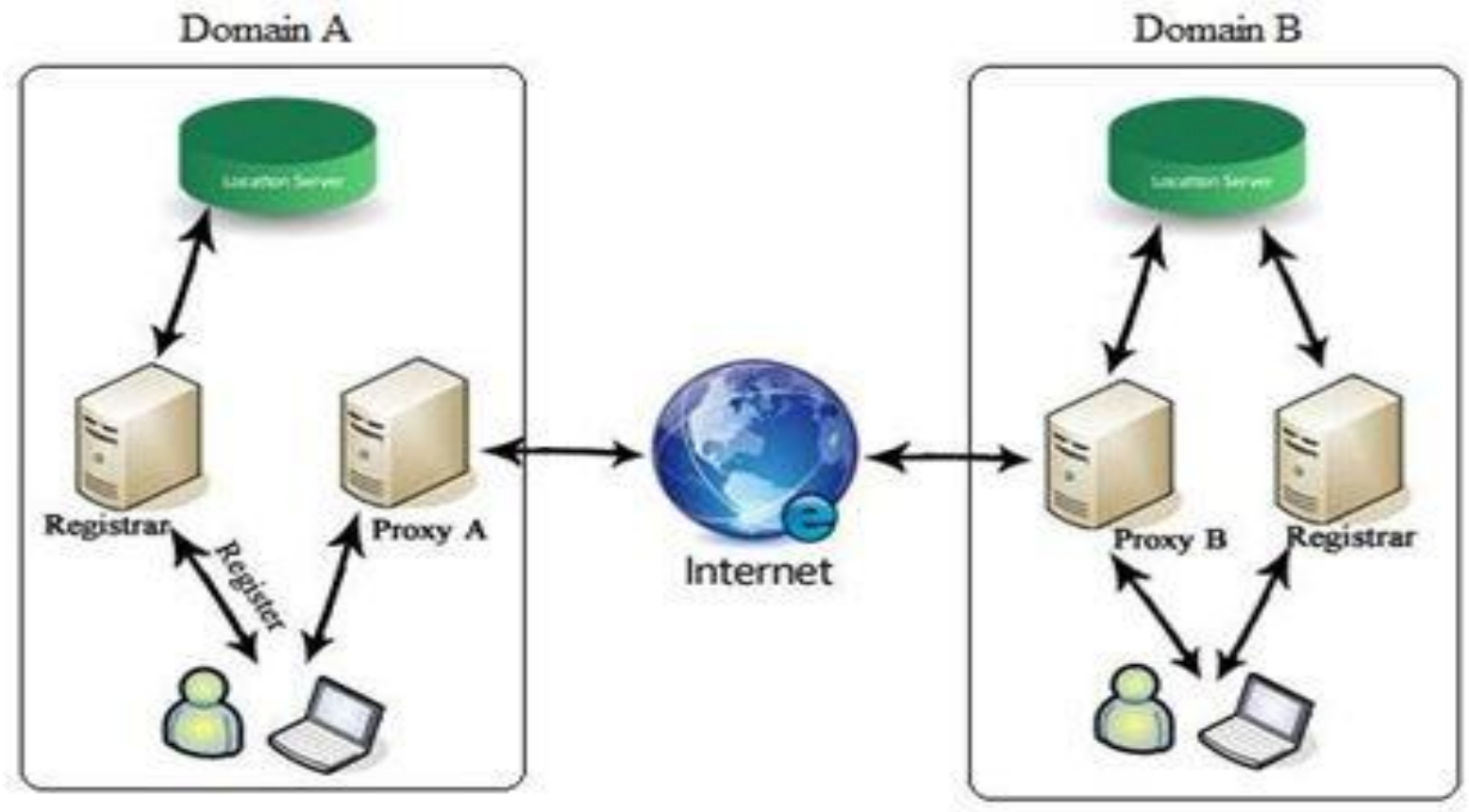

Fig. 2: SIP architecture

## 3. Authentication Scheme

In this section, three authentication schemes by Yang et al. [13], Durlanik et al. [6] and Wu et al. [12] are presented. Yang et al.'s authentication scheme proposed solving the problem of the original SIP mechanism. The original SIP is based on HTTP digest where it is vulnerable to attacks such as off-line password guessing and server spoofing. The improved authentication scheme is based on the Diffie-Hellman Key

Exchange [7], which works based on the difference of Discrete Logarithm Problem (DLP) to resist attacks and increase the security of the SIP authentication scheme. The improved scheme has two phases, namely:

1) Registration Phase: In this phase, if a user (client) wants to register, the person must first submit a request (username and password) to a remote server. When the server receives this user's username and password, it stores them.
2) Authentication Phase: If a legal user wants to login into a system, this user must enter his/her username and password.

These two authentication scheme phases are accomplished through four steps described below and illustrated in figure 3.

*Step1:Client→Server* Client sends a REQUEST that contains {username, $k_1 \oplus F(pw)$ } to the server. User selects $h_1$ randomly and calculates $k_1 = g^{r1} \bmod p$. Note that $f(.)$ is a one-way hash function and $\oplus$ is an operator. User must keep $k_1$ for level three, after that she/he can discard it.

*Step2:Server→Client* Server receives CHALLANGE packet which contains $F(pw) \oplus (k_1 \oplus (pw))$ and then randomly choose $k_2$, $k_2 = g^{r2} \bmod p$, $k = K_1^{r2} \bmod p$, and $F(k_1, k)$. Finally, server sends the CHALLANGE packet to the client.

*Step3:Client→Server* User computes $k = k_2^{r1}$ and $F(k_1, k)$. If the packet is true it sends RESPONSE $(username, realm, F(username, realm, k))$ to the server.

*Step4:Server→Client* When the server receives the RESPONSE packet, it will verify if the packet is true, if it is not, the server will reject it.

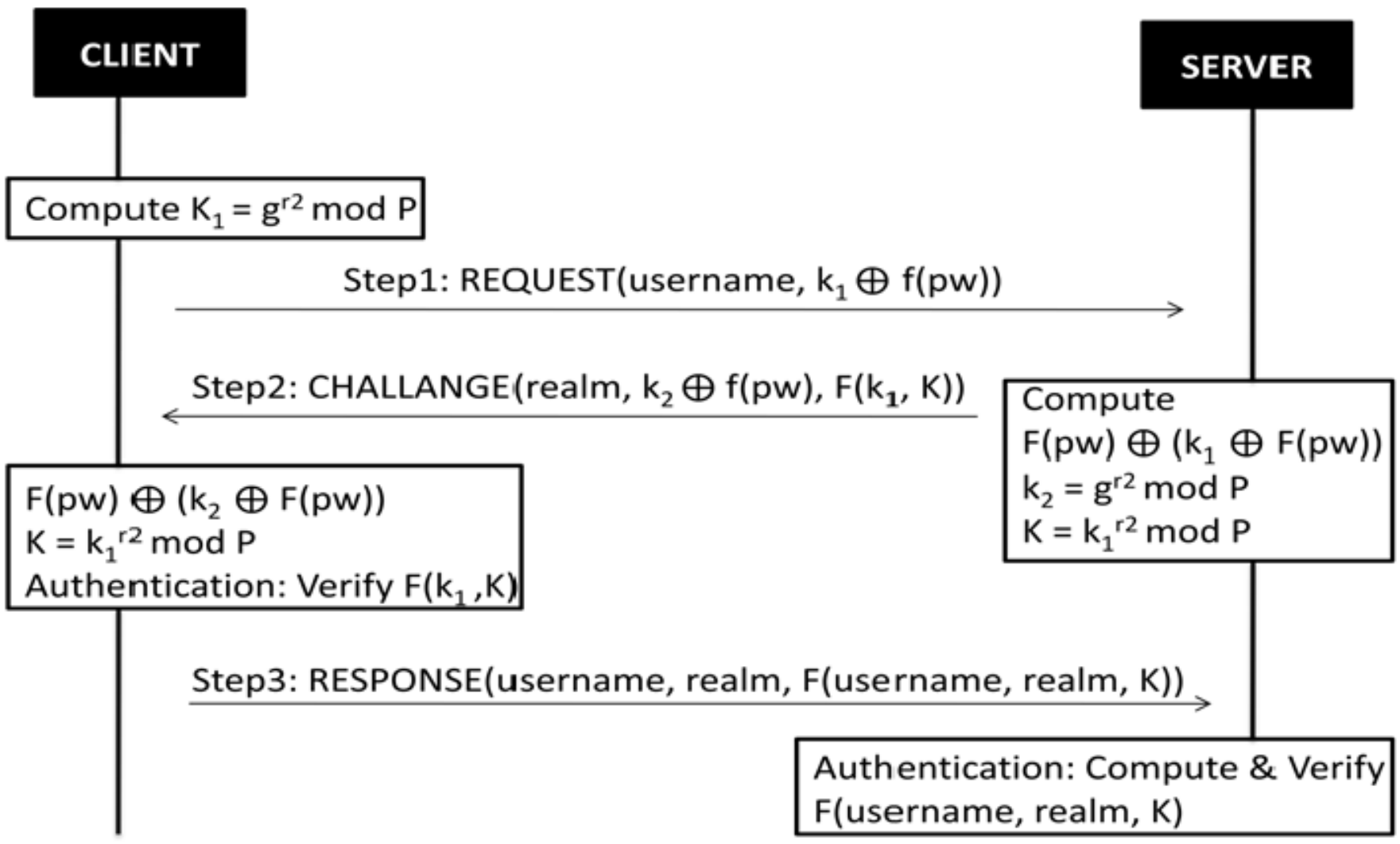

Fig. 3: Secure SIP authentication procedure

A drawback of this authentication scheme is that it is not suitable for devices with low computational power because of the high computational costs. Durlanik et al. [6] proposed an efficient SIP authentication scheme by using an elliptic curve cryptosystem (ECC) [8, 9]. ECC security is based on the elliptic curve discrete logarithm problem (ECDLP). Durlanik et al.'s proposed scheme has both a registration phase and authentication phase. According to Durlanik et al., the authentication scheme requires the server and the user to share a password for authentication which is called a pre-shared password and it uses an elliptic curve public key pair. During the authentication phase, if a legal user wants to login into a system, the user must enter his/her username and password. It offers a level of security comparable to a classical cryptosystem that uses a larger key size. Durlanik et al.'s authentication scheme has some advantages compared to Yang et al.'s, namely that it reduces the total execution time and memory requirements. The attacker does not have the password, $f(pw)$, so she/he is not able to send a RESPONSE message to the server. Hence, a replay attack does not work in this scheme, probably for the following reasons: First, the attacker guesses a password, pw′ randomly and calculates $f(pw')$. Then the attacker calculates these two formulas, (1) and (2):

$$k_1' = F(pw') \oplus (k_1 \oplus F(pw)) \quad (1)$$

$$k_2' = F(pw') \oplus (k_2 \oplus F(pw)) \quad (2)$$

At this stage, the attacker cannot calculate the value of k because she/he may discover the difficulty of a discrete logarithm. Accordingly, the off-line password guessing attack does not work in this scheme. For the following reasons, server spoofing does not work in this scheme either:

$$k_1 = F(pw') \oplus (k_1 \oplus F(pw)) \quad (3)$$

$k = k_1^{h2} \bmod p$ and sends $F(k_1, k)$ to user. User computes $F(pw) \oplus (k_2 \oplus F(pw))$, $k = k_2^{h1} \bmod p$ and verifies $F(k_1, k)$.

## 4. Security Model: Canetti-Krawczyk

Wu et al. discuss a conventional SIP authentication scheme [12] and it shows that the scheme is vulnerable to off-line password guessing attacks [10, 11]. Wu et al. improved the scheme by proposing a new authenticated key exchange protocol called a NAKE. NAKE uses ECC cryptography as a solution to the authentication and key agreement problems that exist in SIP. This mechanism provides mutual authentication and demonstrates the security mechanism in the Canetti-Krawczyk (CK) model. This solution fits efficiently in SIP protocols as described in RFC 3261 [12]. The common parameters used throughout the paper are summarized as:

- $C$ : the client
- $S$ : the Server
- $Pc$ : an identity of Client
- $P_s$ : an identity of Server
- $k$ : a shared secret password between Client and Server
- $realm$ realm: a realm string
- $s$ : an opaque string which is used as a session identifier
- E (Fq): a nonsingular elliptic curve on a finite field Fq
- $P$ : a base point on an elliptic curve ∈ E (Fq) of order n
- $f(.)$ : a secure one-way hash function such as SHA-2
- $\oplus$ : a bit-wise exclusive-or operation

The CK security model presents SK-Security which allows modular design and analysis of key-exchange protocol. It makes difficulty of design and analysis of security protocol simpler. The attacker can have access to three types of secret information: session-start reveal, session-key queries, and party corruption [13].

### 4.1. Security Proof of NAKE protocol

The new SIP authentication mechanism and key agreement protocol proposed here meet the goal and requirements defined. The cryptographic primitive used to provide the assurances are provably secure in the CK security model. Non-repudiation, protection against replay and session hijacking attacks, and mutual authentication are by-products of employing ECC cryptography and hash value.

Replay attack: These attacks, which cannot work in this scheme, can be provided by the freshness of session id s.

Off-line password guessing attack: The attacker guesses a password, k, and computes $f(k)$, after which he/she computes $f(k, Pc, s, (f(k) \oplus (f(k) \oplus \beta)), (f(k) \oplus \alpha)) P_s$ and $f(k, P_s, s(f(k) \oplus \alpha)), (f(k) \oplus (f(k) \oplus \beta)) Pc)$. Obviously, it is not possible for the attacker to compute the value k to match the RESPONSE, because it faces the difficulty of discrete logarithms. Therefore, the protocol is immune to the off-line password guessing attack.

Server spoofing: The server computes shared key $sk = x.\beta$ and sends $f(k, p_s, s, \alpha, \beta, P_c)$ to the client who can then verify the identity of the server by computing $f(k, p_s, s, \alpha, \beta, P_c)$. So, the attacker cannot impersonate the server to deceive the client. Meanwhile, the client derives the shared key by computing $sk = x.\alpha$ and sends $f(k, P_c, s, \alpha, \beta, P_s)$ to the server. Finally, the server can verify the client's identity.

Mutual authentication: Both parties produce a hash value based on a pre-shared key for mutual authentication, and meet the mutual authentication security objectives.

Mutual key agreement and control: Protocol AKE is based on Diffie–Hellman key exchange, the freshness of session key to ensure appropriate selection of random numbers. The two sides have part of a separate key based on a hash value produced by a pre-shared key.

Security parameters are $\alpha$ and $\beta$, each randomly selected by the server and client. Thus, the server and client do not have control of key generation. Fig. 4 shows the improvement of NAKE protocols.

Wu et al. claim that their scheme improves the security attributes required by the SIP standard with minimum changes, and is designed to provide data confidentiality, data integrity, authentication, access control, and perfect forward security. It is secure from any well-known crypto graphical attacks such as replay, off-line password guessing, man-in-the-middle, and server spoofing attacks. Compared to previous schemes [13, 6], Wu et al.'s is more efficient and preferred for applications requiring low memory and rapid transactions.

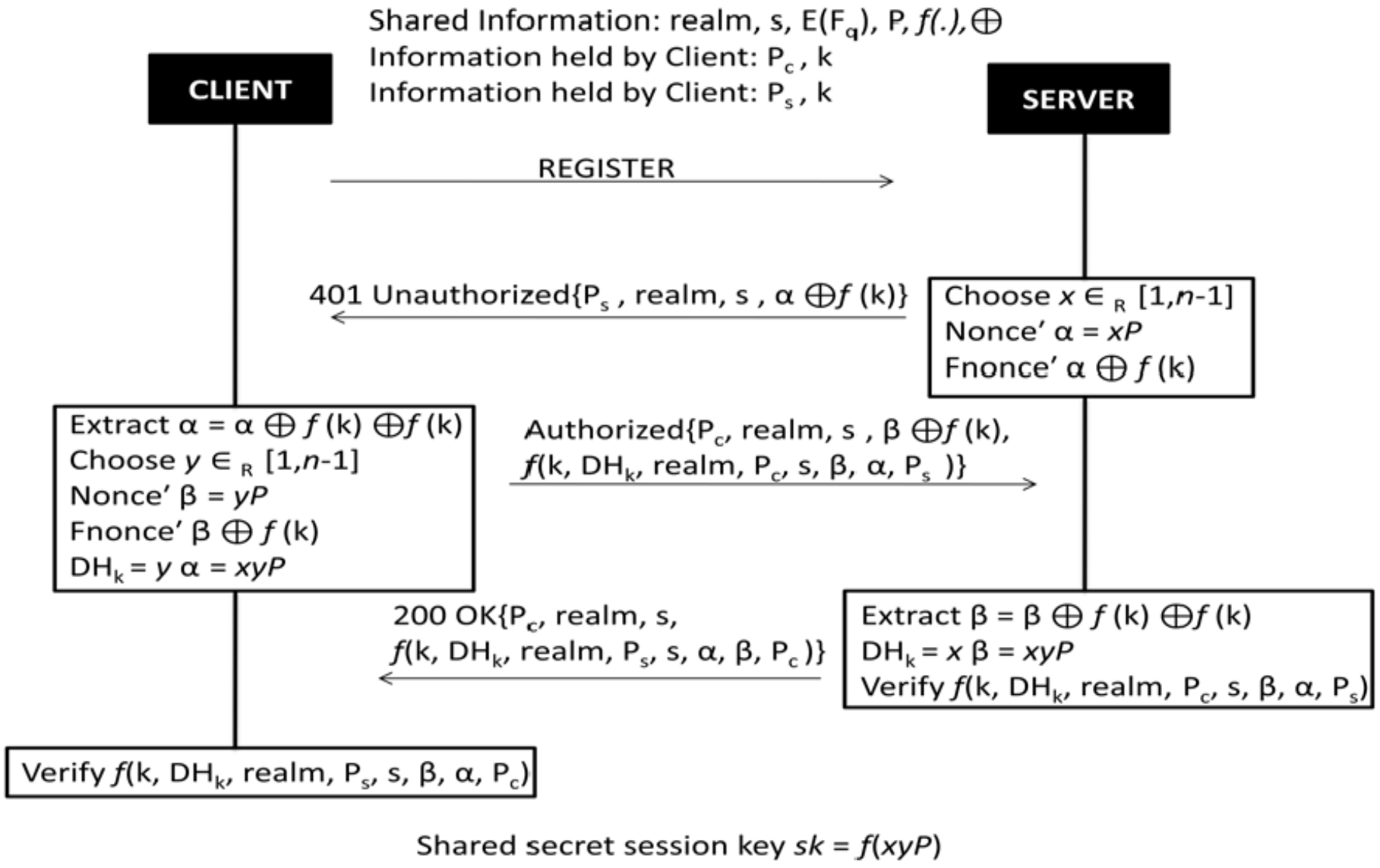

Fig. 4: Nake protocol

## 5. Performance Comparison

The previously presented schemes' computation costs [13, 6, and 12] are shown in table 1. Generally, the Elliptic Curve Discrete Logarithm problem (ECDLP) with an order of 160 bit prime offers approximately the same security level as the Discrete Logarithm Problem (DLP) with 1024 bit modulus. Wu et al.'s authentication scheme requires four exponentiations, four ECC multiplications and six hash operations for protocol execution. Four ECC computations are needed to prevent attacks and to provide perfect forward secrecy. When considering hashing and exclusive-or operations, the proposed scheme requires four hashing operations and four exclusive-or operations for mutual authentication. Obviously, Wu et al.'s scheme is more efficient than others for SIP. The comparison of all the reviewed protocols is summarized against various types of attacks in the figure 5.

Table. 1: Nake protocol

| | DLP | ECDPL | |
|---|---|---|---|
| | [13] | [6] | [12] |
| # OF EXPONENITITIONS | 4 | 0 | 0 |
| # OF ECC COMPUTATIONS | 0 | 4 | 4 |
| # OF HASH FUNCTIONS | 8 | 8 | 6 |
| # OF EXCLUSIVE-OR | 4 | 4 | 4 |
| # OF ROUNDS | 3 | 3 | 3 |

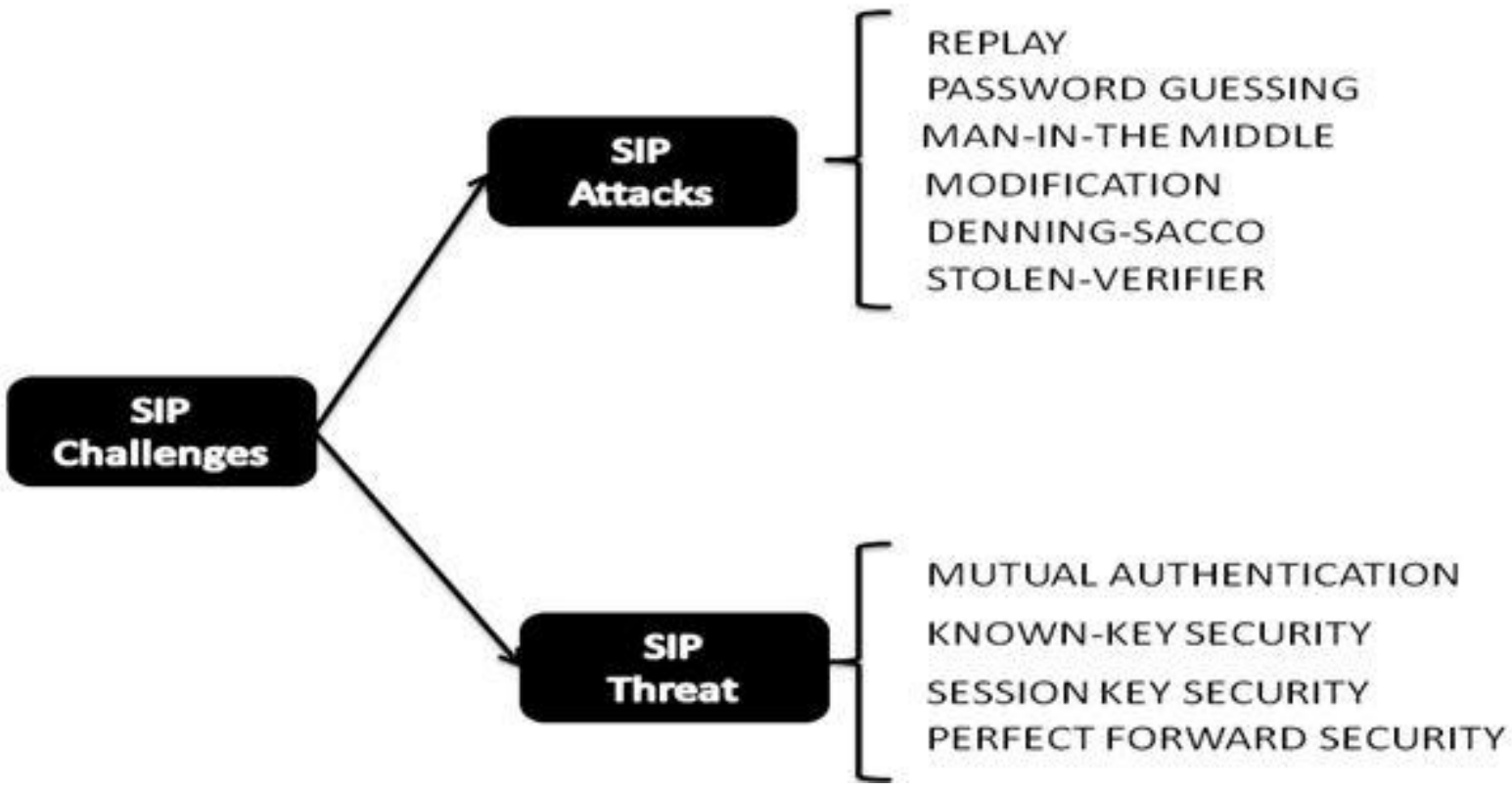

Fig. 5: Taxonomy of SIP threat

Password guessing attack is secure in [13, 6] but the security factor is not considered in [6]. Security policy for Denning-Sacco attack is not provided in [13]. Although in [6] the researcher tried to propose algorithm to tackle the mentioned attack but it does not work properly. Finally, [6] provide algorithm to support security in SIP against this attack. Known-Key, Session key and Prefect and forward security are not available in [13] but [6] had offered a method to handle these threats.

## 6. Conclusion

SIP attacks and authentication schemes have been reviewed to find the best scheme to be implemented in SIP. From the investigation we found that improvement of the NAKE model, which uses ECC cryptography, provides an efficient authentication scheme. Even though this authentication scheme can reduce the risk of attacks, there are some limitations in this scheme's scope such as huge computational load which can be taken into consideration by utilizing fast processing units to alleviate it. The Stolen-Verifier attack can be the pointer for future work.